\documentclass[finallayout,an,fleqn]{w-art}
\newcommand{\sgra}    {\mbox{\rm\,Sgr~A$^*$}}

\usepackage{times}
\usepackage{w-thm}
\theoremstyle{plain}

\theoremstyle{definition}

\usepackage[]{graphicx}
\chardef\bslash=`\\ 

\begin{document}

\DOIsuffix{theDOIsuffix}
\Volume{324}
\Issue{S1}
\Copyrightissue{S1}
\Month{01}
\Year{2003}
\pagespan{3}{}
\Receiveddate{3 March 2003}
\keywords{Accretion, accretion disks --- Black hole physics --- Galaxy:
center --- X-rays: general}
\subjclass[pacs]{04A25}



\title[\sgra\ X-Ray flare seen with XMM-Newton]{A New X-Ray Flare from the Galactic Nucleus
Detected with XMM-Newton}


\author[A. Goldwurm et al.]{A. Goldwurm\footnote{Corresponding
     author: e-mail: {\sf agoldwurm@cea.fr}, Phone: +33\,01\,6908\,2792,
     Fax: +33\,01\,6908\,6577}\inst{1}} \address[\inst{1}]{Service d'Astrophysique, DAPNIA/DSM/CEA, CE-Saclay, F-91191 Gif-Sur-Yvette, France}
\author[]{E. Brion\inst{2}}
\author[]{P. Goldoni\inst{1}}
\author[]{P. Ferrando\inst{1}}
\author[]{F. Daigne\inst{1}}
\author[]{A. Decourchelle\inst{1}}
\author[]{R. S.~Warwick\inst{3}}
\address[\inst{2}]{Centre d'Etude Nucl\'{e}aire de Bordeaux-Gradignan, All\'{e}e du Haut Vigneau, 33175 Gradignan, France}
\address[\inst{3}]{Department of Physics and Astronomy, University of Leicester, Leicester LE1 7RH, UK}
\author[]{P. Predehl\inst{4}}
\address[\inst{4}]{Max-Planck Institut f\"{u}r Extraterrestrische Physik, Postfach 1312, 85741 Garching, Germany}
\begin{abstract}
The compact radio source \sgra\, believed to be the counterpart 
of the massive black hole at the Galactic nucleus, 
was observed to undergo rapid and intense flaring activity in X-rays 
with Chandra in October 2000. 
We report here the detection with XMM-Newton EPIC cameras 
of the early phase of a similar X-ray flare from this source, 
which occurred on 2001 September 4.
The source 2-10~keV luminosity increased by a factor $\approx$~20 
to reach a level of 4~10$^{34}$~erg~s$^{-1}$ in a time interval 
of about 900~s, just before the end of the observation.
The data indicate that the source spectrum was hard 
during the flare and can be described by simple power law of slope  
$\approx$ 0.7.
This XMM-Newton observation confirms the results obtained by Chandra,
suggests that, in \sgra, rapid and intense X-ray flaring is not a rare event
and therefore sets some constraints on the emission mechanism models proposed 
for this source.
\end{abstract}
\maketitle                   





\section{Introduction}

The bright, compact radio source \sgra\ is believed to be 
the radiative counterpart of the 2.6~10$^{6}$~M$_{\odot}$ 
black hole which governs the dynamics 
of the central pc of our Galaxy (Melia $\&$ Falcke 2001).
The compelling evidence for such a massive black hole 
at the Galactic Center (see Sch\"{o}del et al. 2002
for the most rescent results), contrasts remarkably with 
the weak high-energy activity of this object.
In spite of the fact that stellar winds and hot gas 
probably provide enough material for a moderate/low level of accretion, 
the total bolometric luminosity of the source amounts to less 
than 10$^{-6}$ of the estimated accretion power 
(Melia $\&$ Falcke 2001, Goldwurm 2001).
This motivated the development of several black hole accretion 
flow models with low radiative efficiency, some of which have  
also been applied to binary systems, low luminosity nuclei 
of external galaxies and low luminosity active galactic nuclei.
These models include spherical Bondi accretion in conditions
of magnetic field sub-equipartition with a very small 
Keplerian disk located within the inner 50 Schwarzschild radii (R$_S$), 
large hot two-temperature accretion disks dominated by advection (ADAF) 
or non-thermal emission from the base of a jet of relativistic electrons
and pairs, and some other variants or combination of the above models.
However any such model still predicts some level of X-ray emission 
from \sgra\ and determining the properties of such emission
would constrain the theories of accretion and outflows in the 
massive black holes and in general in compact objects.

The 20 years search for high energy emission from \sgra\ 
(Watson et al. 1981, Predehl $\&$ Tr\"{u}mper 1994, Goldwurm et al. 1994) 
has recently come to a turning point with the remarkable observations 
made with the Chandra X-ray Observatory in 1999 and in 2000. 
Baganoff et al. (2001a) first reported the detailed 0.5$''$ 
resolution images obtained
with Chandra in the range 0.5-7~keV, which allowed the detection 
of weak X-ray emission from the radio source. 
The derived luminosity in the 2-10~keV band was 2~10$^{33}$~erg~s$^{-1}$,
for a distance of 8~kpc with steep power law spectrum (index of 2.7) and
some evidence that the source is in part extended on a 1$''$ scale.
Then, in October 2000, the same source was seen to flare up by
a factor of $\approx$~45 in a few hours (Baganoff et al. 2001b).
The luminosity increased from the quiescent level measured in 1999 
to a value of 10$^{35}$~erg~s$^{-1}$. 
The flare lasted a total of 10~ks but the shortest variation 
took place in about 600~s, implying activity on length
scales of $\approx$~20~R$_S$, 
for the above quoted mass of the galactic center black hole.
Evidence of spectral hardening during the flare was also reported by 
the authors who determined a source power law photon index during
the event of 1.3 ($\pm$~0.55), 
significantly flatter than observed during the quiescent state.
These results constrain models of the accretion flow and radiation
mechanism for \sgra.

XMM-Newton, the other large X-ray observatory presently in operation, 
features three large area X-ray telescopes 
coupled to three CCD photon imaging cameras (EPIC) operating 
in the 0.1-15 keV range and to two reflection grating spectrometers 
(RGS) working in the 0.1-2.5 keV band (Jansen et al. 2001). 
Although its angular resolution (6$''$ FWHM) is insufficient for
properly resolving \sgra\ in quiescence, an intense flare such as the 
one seen by Chandra can be easily detected and studied with XMM-Newton.
We report here (see also Goldwurm et al. 2003)
the detection of such en event during a 26 ks XMM-Newton observation
of the Galactic nucleus performed on 2001 September 4 as part of 
a large Galactic Center survey program with XMM-Newton (Warwick et al. 2003).

\section{Results}
The EPIC data reduction and analysis of this XMM-Newton observation are 
described in detail in Goldwurm et al. (2003).
The image recorded in the central CCD (11$'$~$\times$~11$'$ for the MOS) 
was dominated by the diffuse emission of the Sgr A East region, and 
in order to search for a variable central source
we extracted and analyzed light curves from events collected within 
a 10$''$ radius region centered on \sgra.
As shown in Fig.~1, the 2-10 keV count rate 
from the combined MOS~1 and MOS~2 events selected in this way,
is quite stable around an average value of 0.08~cts~s$^{-1}$ till the last
900~s of the observation.
Then the count rate gradually increases to reach a value of about a factor 
3 higher in the last bin. 
The integrated count rate in the last 900~s reaches 7~$\sigma$ over 
the average value measured before the flare and the detected variation 
has a negligible probability to be a statistical fluctuation.
A similar peak  (4.3~$\sigma$) is observed in the counts extracted from the 
PN camera which stopped observing about 250~s before the MOS (see Fig. 1 b)). 
Similar light curves, from a larger region far from the source 
do not show any evidence of such an increase in the counting rate.

In Fig.~2 we report a MOS image of the region around the nucleus
integrated during the 1000~s before the flare and a similar image 
integrated during the last 1000~s and fully including the source flare.
The brightening we detected in the light curves is clearly due to 
the brightening of a central source.
We compared data to the instrument point spread function to determine 
the location of the excess. 
On the 2-10 keV MOS 1 and MOS 2 image of the last 1000~s, rebinned 
to have pixel size of 4$''$, we obtained the centroid of the source at 
R.A. (J2000) = 17$^h$~45$^m$~39.99$^s$ 
~Dec (J2000) = -29$^{\circ}$~00$'$~26.7$''$,
with a total error, dominated by residual systematic uncertainties
in the XMM-Newton focal plane, of 1.5$''$.
The derived flare position is therefore compatible with \sgra\ 
radio location (Yusef-Zadeh et al. 1999), since it is offset from the latter 
by only 1.5$''$ and we conclude that 
the flare detected by XMM-Newton is associated with the galactic nucleus.

A first spectral analysis of the flaring event was performed by computing 
a simple hardness ratio, 
defined as the ratio between the measured counts (including background) 
in the hard band 4.5-10 keV, and those in the soft band 2-4.5 keV. 
We found that the hardness ratio increased by 0.32~$\pm$~0.13 
during the event with respect to the value before the flare.
Though the hardening has a modest statistical significance 
of 2.5~$\sigma$, 
it is fairly consistent with the flare trend observed with Chandra.
Since Chandra data showed that the quiescent emission within
10$''$ from \sgra\ position contains a dominant diffuse emission and
a large contribution from close point sources, we had to model
in someway this components to study the flare spectrum
(see Goldwurm et al. 2003).
We extracted MOS and PN count spectra from the 10$''$ radius
circular region centered on \sgra\ before the flare and
during the flare (last 900~s for MOS and last 700~s for PN).
To derive the spectra reported in Fig. 3 we used the spectra extracted
before the flare as background components for the flare spectra.
After subtraction of the non-flaring count spectrum, the flaring 
MOS and PN spectra were fitted with a simple absorbed power law with N$_H$ 
fixed to the Chandra measured value of 9.8 10$^{22}$ cm$^{-2}$ 
and leaving untied the MOS and PN normalizations. 
Results both without and including dust scattering are reported in Table 1.
We obtained a best fit photon index of 0.7~$\pm$~0.5 
(error at 1 $\sigma$ for one interesting parameter) with $\chi_{\nu}^2$~=~0.98
for 20 d.o.f., that is significantly harder than
the spectrum measured with Chandra during the quiescent state 
(2.7~$\pm$~1.0), and compatible, within uncertainties, to the
index measured during the 2000 October flare.
This procedure subtracts from the flare spectrum the 
non flaring component of \sgra\
and therefore assumes that the quiescent emission from \sgra\
is negligible.
This is an acceptable approximation since, if at the level
measured in 1999 by Chandra, the quiescent emission 
contributes by only $\approx$~5$\%$ to the counts of the flare spectrum. 
On the other hand, this procedure allows to subtract
the diffuse emission present in the region of the spectral extraction 
and the instrumental background in a model-independent way.
We remark that the count excess around 6-7 keV in the residuals
of the MOS spectrum of Fig. 3 is not significant. 
Including a narrow gaussian line at 6.4 keV (with fixed centroid and
zero width) to the model of the absorbed power law, 
we can set an upper limit (90~$\%$ confidence level in 1 parameter) 
to an iron emission line of about 1.8 keV equivalent width.

The measured absorbed source flux in the 2-10 keV band, 
corrected for the fraction of encircled energy at a distance of 10$''$ 
(60$\%$), is then of (3.3~$\pm$~0.6)~10$^{-12}$~ergs~cm$^{-2}$~s$^{-1}$
(1~$\sigma$ errors by fixing best fit parameters but normalization), 
equivalent to a 2-10~keV luminosity at 8~kpc of $\approx$~4~10$^{34}$~erg~s$^{-1}$.
This is the average value in the last 900~s but the last light
curve 180~s bin was about a factor 1.4 higher, thus the luminosity
reached a value of $\approx$~6~10$^{34}$ erg~s$^{-1}$.
These numbers are subject to large errors due to the low
statistics available. But the general result which emerges is that 
the flare we detected presents a harder spectrum than the one
measured with Chandra for \sgra\ during the quiescent period.

\begin{figure}[htb]
\includegraphics[width=.45\textwidth]{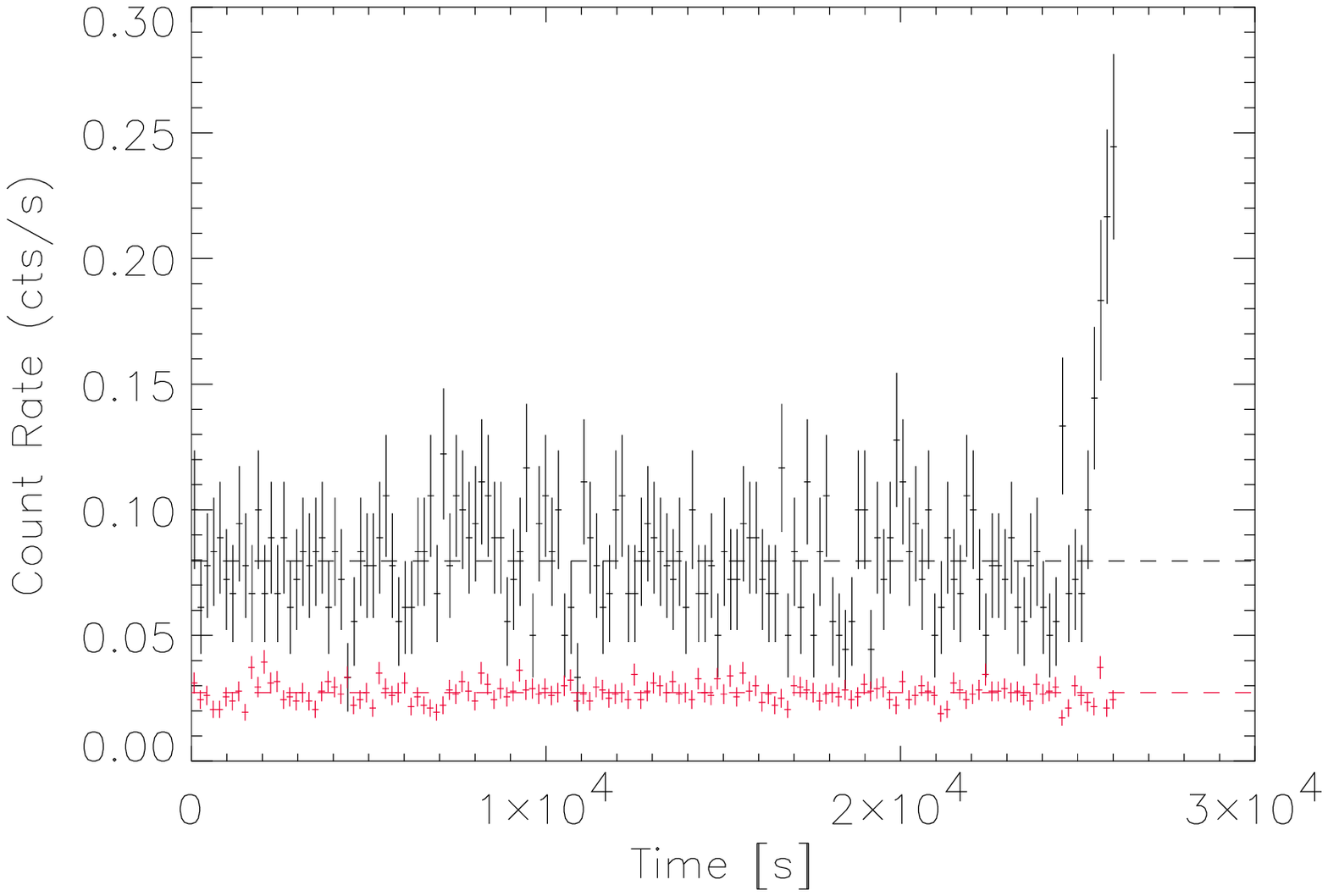}~a)
\hfil
\includegraphics[width=.45\textwidth]{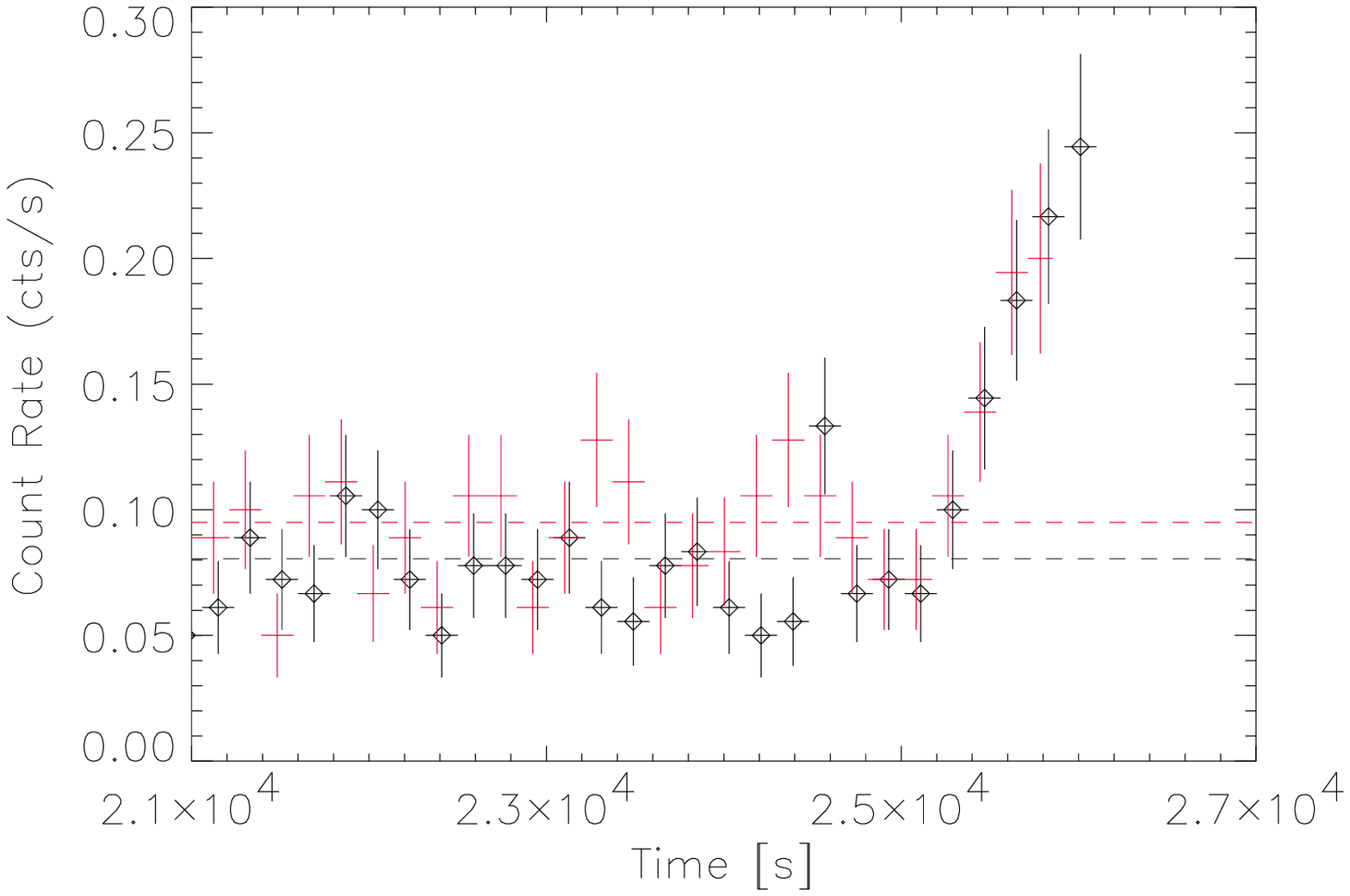}~b)
\caption{a) Count rate, sampled in bins of 180~s, collected with both MOS 
cameras from a region within 10$''$ from \sgra\ in the range 2-10 keV
(black upper curve). An equivalent light curve collected from a 30$''$ radius
region centered about 1$'$ East of \sgra\ and rescaled by a factor 0.1
for clarity, is shown for comparison (red lower curve).
Dashed lines indicate the average value computed before the flare.
b) Zoom of the \sgra\ MOS light curve (black circles) around 
the period of the flare compared to a similar light curve (count rate
within 10$''$ from \sgra\ in the 2-10 keV band in bins of 180~s) 
from PN data (red crosses)}
\label{fig:1}

\end{figure}
\begin{figure}[htb]
\includegraphics[width=.45\textwidth]{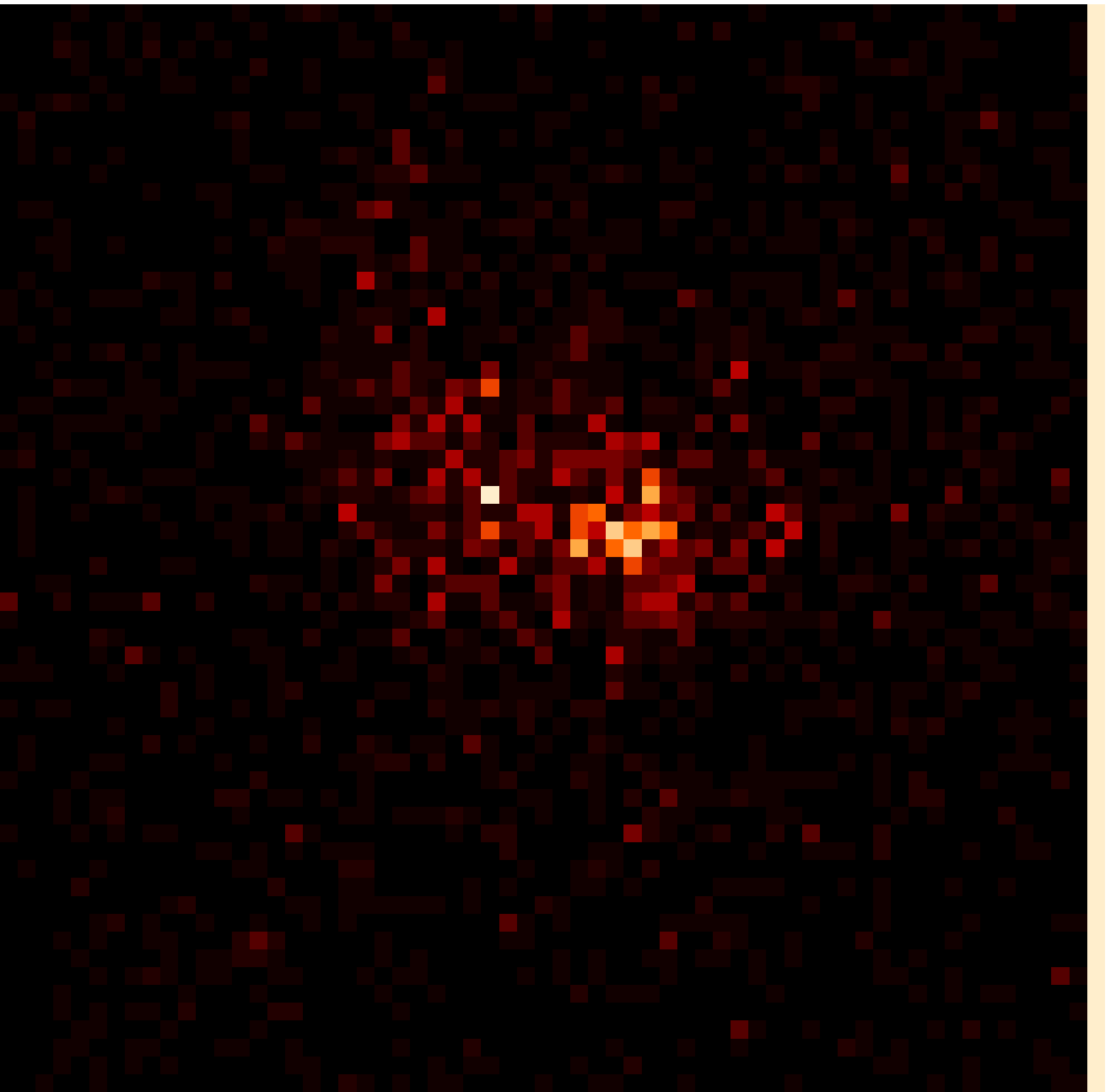}~a)
\hfil
\includegraphics[width=.45\textwidth]{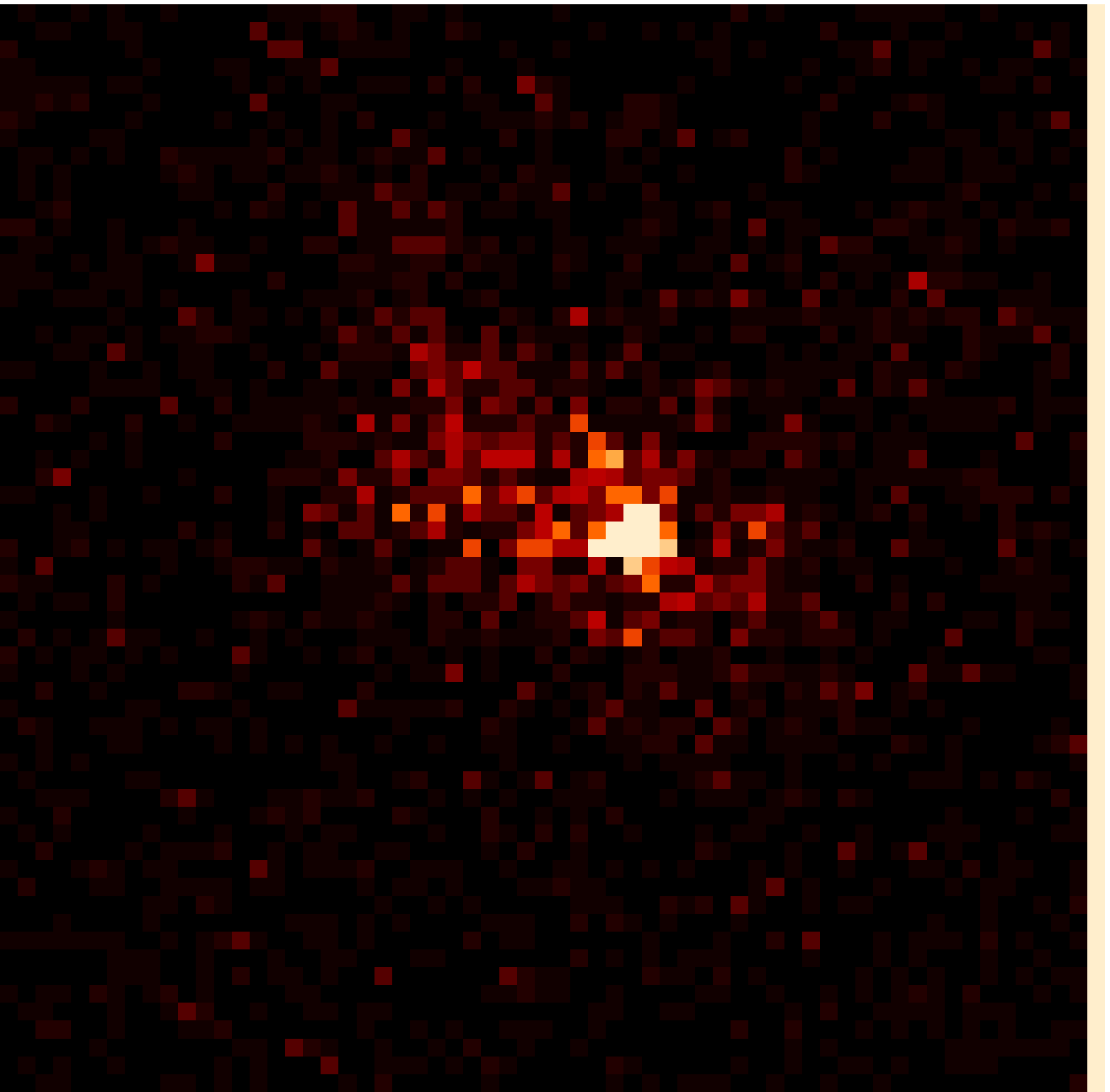}~b)
\caption{Images of the 5$'$~$\times$~5$'$ region around the Galactic
nucleus in the band 2-10 keV obtained from MOS events integrated
in the 1000~s before the flare (a) and in the last 1000~s of the 
observation including the flare (b). 
Pixels were rebinned to a size of 5.5$''$~$\times$~5.5$''$. 
\sgra\ position is right in the middle of the central bright pixel
visible in the flare image (b).}
\label{fig:2}
\end{figure}

\section{Discussion}

The XMM-Newton discovery of a new X-ray flare of \sgra\ in September 2001
confirms the results obtained in the earlier Chandra observations.
XMM-Newton observed only the first part of the flare, 
but the recorded event is fully compatible in intensity, spectrum and time 
scales with the early phase of the flare seen by Chandra.
This detection of another such large X-ray flare from \sgra\ indicates that 
the event is not rare. In spite of the lack of detection of flares
from \sgra\ in another 50 ks XMM-Newton observation performed in February 2002 
(Predehl et al. 2003), the dayly presence of X-ray flares 
in \sgra\ has been recently confirmed by further Chandra observations
performed in 2002 (Baganoff et al. 2003).
The radio source on the other hand has been observed many times 
and the detected flux variability has never exceeded a factor 2 
(Zhao et al. 2001).
This implies that it is unlikely that radio emission
presents a comparable large increase in flux and this provides some constraints 
on the models.

The X-ray flare from \sgra\ cannot be explained by pure Bondi or
ADAF models (Narayan et al. 1998) as in these models the emission 
is due to thermal bremsstrahlung
from the whole accretion flow and arises from an extended region 
(between 10$^3$ - 10$^5$ R$_S$) which cannot account for such 
rapid variability. 
Models which predict emission from the innermost regions
near the black hole involve a mechanism acting either at the base 
of a jet of relativistic particles (Markoff et al. 2001)
or in the hot Keplerian flow present within the circularization radius
of a spherical flow (Melia et al. 2001, Liu $\&$ Melia 2002).
In both cases a magnetic field is present in the flow and 
the linearly polarized sub-mm radiation is attributed to optically thin 
synchrotron emission from the inner region, while the X-rays 
during quiescent period are produced 
by the synchrotron self-Compton (SSC) mechanism whereby radio to mm
photons are boosted to X-ray energies by the same relativistic 
or subrelativistic electrons that are producing the synchrotron radiation.
However large X-ray flux variations produced by a change in accretion rate 
in these models would imply an equivalent increase in the radio and sub-mm 
part of the spectrum
not compatible with the lower amplitude of radio changes compared 
to X-rays (Markoff et al. 2001). Not to mention that the X-ray spectrum would 
remain rather steep.
The model of a inner circularized flow however predicts 
low or anti correlation of the radio emission with the X-rays
if the radiation mechanism for the X-ray flare is bremsstrahlung rather than SSC.
The sub-mm and far IR domain on the other hand
would in this case show a large correlated increase, 
but at these frequencies the measurements have not been frequent 
enough to settle the issue.
Though the exact modelling of radiation process depends 
on viscosity behavior and other uncertain details, 
the observed hardening of the spectrum during the flare indeed
favours the bremsstrahlung emission mechanism in this model rather 
than the SSC one (Liu $\&$ Melia 2002).
Another totally different model for the X-ray flares 
(Nayakshin $\&$ Sunyaev 2003) proposes that those result from
the interraction of close orbiting stars with a very cold 
neutral accretion disk around \sgra.

More compelling constraints on the models will be set when 
simultaneous observations in radio/sub-mm and X-ray wavelengths
of such a flare are obtained.
We compared the time of the flare to a recent radio light curve
of \sgra\ obtained at 1.3 cm and 2 cm with the VLA between 2001 March  
and November (Yuan $\&$ Zhao 2003). The X-Ray flare 
occurred 1-2 days after a local maximum of the curve, but no radio data
points are reported for the day when our XMM-Newton observation
took place.
It will be also important to study the shape of the flare spectrum at 
energies higher than 10 keV to fully understand the radiation mechanism 
producing the X-rays.
In particular by measuring the high energy cut-off of the spectrum  
one could determine the electron temperature for a thermal emission 
or the Lorentz factor for non-thermal processes.
We estimated that such a flare should be marginally visible in the 
range 10-60 keV
with the low energy instruments on board the new gamma-ray mission 
INTEGRAL, operating since 2002 October,  
if the spectrum extends to these energies with the slope observed 
with Chandra and XMM-Newton.

\begin{acknowledgement}

This paper is based on observations with \textit{XMM-Newton}, an ESA science mission 
with instruments and contributions funded by ESA member states and the USA 
(NASA).
\end{acknowledgement}

\begin{vchtable}[htb]
\vchcaption{Spectral Fit to X-ray Emission from within 10$''$
   from \sgra\ during the Flare}
\label{tab:3}\renewcommand{\arraystretch}{1.5}
\begin{tabular}{lcc} \hline

Power-law Model & No Dust Scattering & Dust Scattering \\ \hline
$\rm N_H$ [$10^{22}$~cm$^{-2}$] & 9.8 & 5.3 \\
Photon Index & 0.7~$^{+0.5} _{-0.6}$ & 0.3~$^{+0.6} _{-0.4}$ \\
Norm MOS [$10^{-4}$ ph cm$^{-2}$ s$^{-1}$ keV$^{-1}$]  & 1.3~$^{+2.0} _{-1.3}$ & 0.7~$^{+1.0} _{-0.4}$ \\
Norm PN [$10^{-4}$ ph cm$^{-2}$ s$^{-1}$ keV$^{-1}$] & 0.3~$^{+0.6} _{-0.3}$ & 0.2~$^{+0.3} _{-0.1}$ \\
$\rm \chi^2_{\nu}~(d.o.f.)$ & 0.98 (20) & 0.95 (20) \\ \hline

\end{tabular}
\\
Notes : \\
Scattering computed for fixed value of A$_{V}$~=~30. \\
Normalization is the flux density at 1~keV. \\
Errors are at 68.3$\%$ confidence interval 
for 1 interesting parameter.
\end{vchtable}

\begin{figure}[htb]
\includegraphics[width=\textwidth, height=9cm]{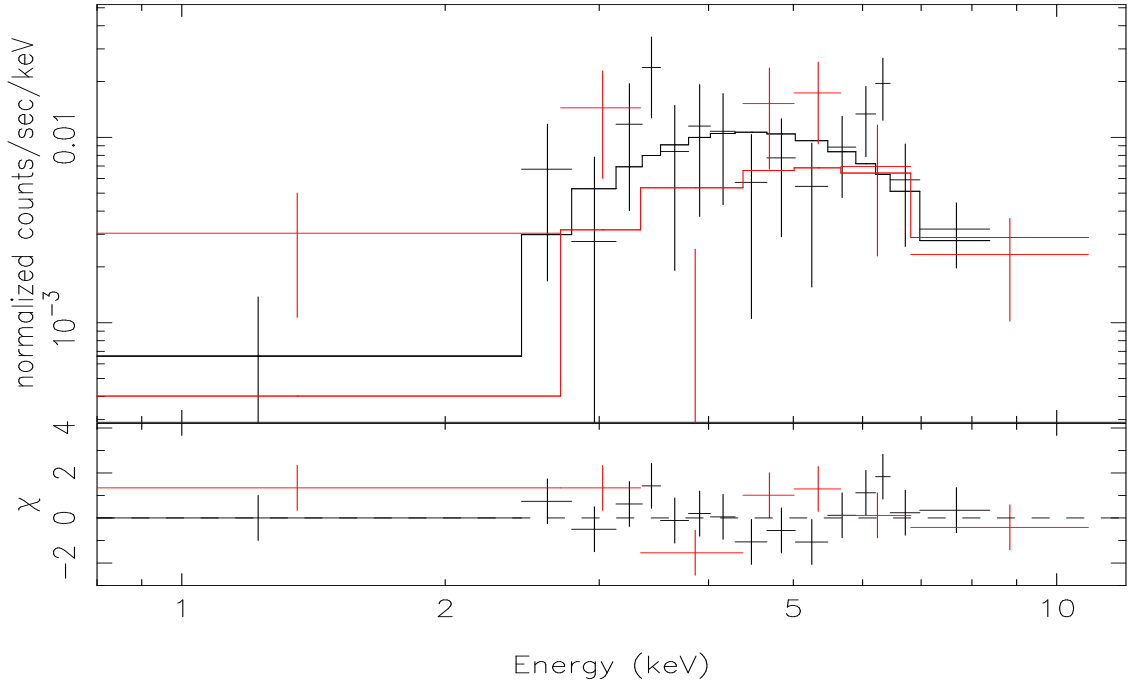}
\caption{Count spectra from MOS (black data point set) and 
PN (red data point set) data, extracted from a region of 10$''$ radius
around \sgra\ during the flare after subtraction of the non flaring 
spectra. 
The spectra are compared to the best fit model of an absorbed power law 
without dust scattering (see Table 1)}
\label{fig:3}
\end{figure}

\end{document}